# Structure of Higher Order Corrections to the Lipatov Asymptotic Form

## I. M. Suslov

*Kapitza Institute of Physical Problems, Russian Academy of Sciences, Moscow, 117334 Russia*
*e-mail: suslov@kapitza.ras.ru*
Received November 1, 1999

**Abstract**—High orders of perturbation theory can be calculated by the Lipatov method, whereby they are determined by saddle-point configurations, or instantons, of the corresponding functional integrals. For most field theories, the Lipatov asymptotic form has the functional form $ca^N\Gamma(N + b)$ ($N$ is the order of perturbation theory) and the relative corrections to it are series in powers of $1/N$. It is shown that this series diverges factorially and its high-order coefficients can be calculated using a procedure similar to the Lipatov one: the $K$th expansion coefficient has the form $\text{const}[\ln(S_1/S_0)]^{-K}\Gamma(K + (r_1 - r_0)/2)$, where $S_0$ and $S_1$ are the values of the action for the first and second instantons of this particular field theory, and $r_0$ and $r_1$ are the corresponding number of zeroth-order modes; the instantons satisfy the same equation as in the Lipatov method and are assumed to be renumbered in order of their increasing action. This result is universal and is valid in any field theory for which the Lipatov asymptotic form is as specified above. © 2000 MAIK "Nauka/Interperiodica".

## 1. INTRODUCTION

Lipatov proposed a general method of calculating high orders of perturbation theory whereby these are determined by saddle-point configurations, or instantons, of the corresponding functional integrals [1]. On its appearance, the Lipatov method stimulated major discussion (see the collection of articles [2]) but was subsequently cast into doubt because of the possible existence of additional renormalon contributions [3]. In a recent paper [4], the present author put forward a detailed discussion of the existing argumentation in support of renormalons and showed that this is untenable in the broad philosophical sense and in the mathematical sense: this clears any obstacles from applying the Lipatov method to a wide range of problems in theoretical physics.

The Lipatov method can be used to study any quantities [5] but the starting point is that it can be applied to functional integrals having the form

$$I(g) = \int D\varphi \varphi^{(1)} \cdots \varphi^{(M)}$$
$$\times \exp(-S_0\{\varphi\} - gS_{int}\{\varphi\}), \tag{1}$$

where $\varphi^{(1)}, \ldots, \varphi^{(M)}$ is a certain sample $\varphi_{i_1}, \ldots, \varphi_{i_M}$ from the set of integration variables $\varphi_i$ contained within the symbol $D\varphi$. The expansion coefficients $I_N$ of the integral (1) in terms of the coupling constant $g$ are determined by the Cauchy integral

$$I_N = \oint_C \frac{dg}{2\pi i} \frac{I(g)}{g^{N+1}}, \tag{2}$$

in which the saddle-point method can be used for large $N$. The functional form of the Lipatov asymptote is given by

$$I_N = ca^N\Gamma(N + b), \quad N \longrightarrow \infty, \tag{3}$$

and the relative corrections to it have the form of a regular expansion in terms of $1/N$:

$$I_N = ca^N\Gamma(N + b)$$
$$\times \left\{ 1 + \frac{A_1}{N} + \frac{A_2}{N^2} + \ldots + \frac{A_K}{N^K} + \ldots \right\}. \tag{4}$$

Calculation of the corrections to the asymptotic form provides important information on the expansion coefficients and is an alternative to the direct diagrammatic calculations of the lower orders: for instance, instead of calculating the fourth or fifth orders [6, 7], it is more economical to calculate $A_1$ or $A_2$. So far, the first corrections to the asymptotic form have only been calculated in $\varphi^4$ theory [8] and in a few quantum-mechanical problems [9, 10].

In the present paper, we study the behavior of the coefficients $A_K$ for large $K$. This topic has not been studied theoretically and the only available data has been obtained by numerical methods: for a perturbation theory series in the problem of an anharmonic oscillator, Bender and Wu [9] determined the first ten coefficients $A_K$:

$$A_1 = -1.3194444, \quad A_3 = -7.0142876,$$
$$A_2 = -1.9385609, \quad A_4 = -40.118943,$$

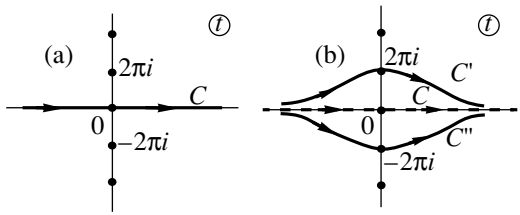

**Fig. 1.** (a) Saddle points and integration contour in integral (11). (b) In calculations of the asymptotic form of $A_K$ the contour must be deformed since the point $t = 0$ corresponds to a singularity not a saddle point.

$$A_5 = -305.5223, \qquad A_8 = -3.65 \times 10^5, \qquad (5)$$

$$A_6 = -2808.09, \qquad A_9 = -4.4 \times 10^6,$$

$$A_7 = -2.995 \times 10^4, \qquad A_{10} = -1 \times 10^8.$$

The rapid growth of these coefficients indicates that the series in (4) diverges.

Another example which can easily be studied is the zero-dimensional limit of $\varphi^4$ theory. In this case, the functional integral in fact reduces to a single one

$$I(g) = \int_0^\infty d\varphi \, \varphi^M \exp(-\varphi^2 - g\varphi^4), \qquad (6)$$

and its expansion coefficients can be calculated in the explicit form:

$$I_N = \frac{2^{M/2}}{2\sqrt{2\pi}} \frac{\Gamma\left(N + \frac{M+3}{4}\right)\Gamma\left(N + \frac{M+1}{4}\right)}{\Gamma(N+1)} (-4)^N. \quad (7)$$

By isolating the asymptotic form for $N \longrightarrow \infty$ and expressing the result in the form (4), we obtain the following expression for the coefficients $A_K$ for $K \longrightarrow \infty$ (see Appendix)

$$A_K = \mathrm{Re}\,\frac{2(1 + e^{\pi i M})}{(2\pi i)^{K+1}}\Gamma(K), \qquad (8)$$

whose functional form is similar to the Lipatov asymptotic form (3) but with the complex parameters $a$ and $c$.

In the present study, we shall show that factorial divergence of the series in (4) also occurs in the general case and a universal result [see formula (47)] valid for any field theory using the Lipatov asymptotic form (3) can be obtained for the asymptotic form $A_K$.

## 2. SIMPLE EXAMPLE AND QUALITATIVE PICTURE

The qualitative aspects involved in the calculation of the asymptotic form of $A_K$ can be conveniently demon-

strated by calculating the corrections to the Stirling formula:

$$\Gamma(N+1) = \sqrt{2\pi N}\, e^{-N} N^N$$

$$\times \left\{ 1 + \frac{A_1}{N} + \frac{A_2}{N^2} + \dots + \frac{A_K}{N^K} + \dots \right\}. \qquad (9)$$

The result is well known for the logarithmic form of expression (9): in this case it is possible to find a general term of the series known as the Stirling series [11]. By calculating the exponential function of the Stirling series using factorial series algebra [5], we can easily find the asymptotic form of $A_K$:

$$A_K = -\mathrm{Re}\,\frac{2\Gamma(K)}{(2\pi i)^{K+1}}, \quad K \longrightarrow \infty. \qquad (10)$$

We shall subsequently show how this result is obtained.

Using the definition of a gamma function and making the substitutions $x \longrightarrow Nx$ and $t = \ln x$, we have

$$\Gamma(N+1) = \int_0^\infty dx \, x^N e^{-x} = N e^{-N} N^N$$

$$\times \int_0^\infty dx \exp\{-N[x - 1 - \ln x]\} \qquad (11)$$

$$= N e^{-N} N^N \int_{-\infty}^\infty dt \, e^t \exp\{-N[e^t - 1 - t]\}.$$

For large $N$, the saddle-point condition has the form $e^t - 1 = 0$ so that there is a set of saddle points $t_s = 2\pi i s$, $s = 0, \pm 1, \pm 2, \dots$, lying on an imaginary axis (Fig. 1a). The integration contour in (11) passes through the saddle point $t = 0$ and satisfies all the conditions for validity of the saddle-point method [12]. Thus, its deformation is not required and the other saddle points can be neglected. Calculating the integral in the saddle-point approximation yields the Stirling's formula.

Formally isolating the asymptotic form, we identically set

$$\Gamma(N+1) = \sqrt{2\pi N}\, e^{-N} N^N F(1/N) \qquad (12)$$

and making the substitution

$$\epsilon = 1/N, \qquad (13)$$

we have for the function $F(\epsilon)$ introduced by us

$$F(\epsilon) = \frac{1}{\sqrt{2\pi\epsilon}} \int_{-\infty}^\infty dt \, e^t \exp\left\{ -\frac{e^t - 1 - t}{\epsilon} \right\}. \qquad (14)$$

Expanding (14) as a series in terms of $\epsilon$ gives the required coefficients $A_K$ which are calculated by analogy with (2):

$$
A_{K-1} = \oint_C \frac{d\epsilon}{2\pi i} \frac{1}{\sqrt{2\pi\epsilon}}
$$

$$
\times \int_{-\infty}^{\infty} dt\, e^t \exp\left\{-\frac{e^t - 1 - t}{\epsilon} - K\ln\epsilon\right\}.
\tag{15}
$$

This is an exact expression which for large $K$ can be calculated using the saddle-point approximation. The saddle-point conditions yield the set of solutions

$$
t_s = 2\pi i s, \quad \epsilon_s = -\frac{t_s}{K}, \quad s = \text{integer},
\tag{16}
$$

so that in the complex plane $t$ the saddle points are formally the same as those in the calculations of the leading asymptotic term. However, for the integrand at the $s$th saddle point we can easily obtain the estimate

$$
\sim \exp\{-K - K\ln\epsilon_s\} \sim t_s^{-K} K!,
\tag{17}
$$

from which it is clear that the solution with $s = 0$ does not in fact correspond to a saddle point but to a singularity.[1] Hence, the integration contour over $t$ cannot pass through the point $t = 0$ but must be deformed and pass through one of the neighboring saddle points, $2\pi i$ or $-2\pi i$ (Fig. 1b) which, because of (17), gives the required asymptotic form of $A_K \sim (2\pi)^{-K} K!$ [see (10)].[2]

A similar situation is encountered in the general case. When the coefficients $A_K$ are calculated in the saddle-point approximation, the instanton equation is the same as that used to calculate the Lipatov asymptotic form. However, using the same solution as in the last case yields a singularity rather than a saddle point (because of the additional integration over $\epsilon$). There is thus a need to consider other solutions of the instanton equation which can be numbered in order of increasing action corresponding to them. If the Lipatov asymptotic form is determined by the first instanton, having the smallest action, the principal contribution to the asymptotic form of $A_K$ is made by the second instanton.

## 3. GENERAL CASE

Calculations of the Lipatov asymptotic form (3) are fairly cumbersome if the aim is to find all its parameters $a$, $b$, and $c$. However, if the analysis is confined to deter-mining the parameters $a$ and $b$, it is possible to have simple structural calculations which reduce to a formal expansion near the saddle point and isolate the dependence on $N$. We shall demonstrate these calculations for the case of $\varphi^4$ theory; however, we do not need the explicit form of the action and we shall only use its characteristic properties of homogeneity

$$
S_0(\lambda\varphi) = \lambda^2 S_0(\varphi), \quad S_{int}(\lambda\varphi) = \lambda^4 S_{int}(\varphi).
\tag{18}
$$

Similar properties of homogeneity are encountered in other field theories and, with slight modifications, the scheme put forward subsequently also holds for the general case.

According to (1) and (2), the expansion coefficients are given by

$$
I_{N-1} = \oint_C \frac{dg}{2\pi i} \int D\varphi\, \varphi^{(1)} \ldots \varphi^{(M)}
$$

$$
\times \exp(-S_0\{\varphi\} - gS_{int}\{\varphi\} - N\ln g).
\tag{19}
$$

We introduce the new variable

$$
\phi = \varphi\sqrt{g}
\tag{20}
$$

and set

$$
S\{\varphi\} = S_0\{\varphi\} + S_{int}\{\phi\}.
\tag{21}
$$

In terms of the new variable $\phi$, the saddle-point conditions have the form

$$
S'\{\phi_c\} = 0, \quad g_c = \frac{S\{\phi_c\}}{N},
\tag{22}
$$

and expanding the expression in the exponential function (19) as far as quadratic terms in $\delta\phi = \phi - \phi_c$ and $\delta g = g - g_c$ gives

$$
-N - N\ln g_c - \frac{N(\delta\phi, S''\{\phi_c\}\delta\phi)}{2\,S\{\phi_c\}} - \frac{N}{2g_c^2}(\delta g)^2.
\tag{23}
$$

We use a symbolic notation, denoting the first and second functional derivatives by single and double primes and taking these to be a vector and a linear operator, respectively; the variables of integration $\varphi_i$ contained within $D\varphi$ are taken to be components of the vector $\varphi$. Bearing in mind that because of (20)

$$
\delta\phi = \sqrt{g_c}\left(\delta\phi + \frac{\delta g}{2g_c}\varphi_c\right), \quad \delta\phi = \phi - \varphi_c,
\tag{24}
$$

and shifting the origin $\delta\phi$, we have

$$
I_{N-1} = e^{-N} g_c^{-N+1-M/2} \int_{-\infty}^{\infty} \frac{dt}{2\pi} \int D\varphi\, \phi_c^{(1)} \ldots \phi_c^{(M)}
$$

$$
\times \exp\left(-\frac{1}{2}(\delta\phi, S''\{\phi_c\}\delta\phi) + \frac{N}{2}t^2\right),
\tag{25}
$$

where we have set $\delta g = ig_c t$.

---

[1] The solutions (16) are written assuming $\epsilon \neq 0$ which does not hold for $s = 0$. A similar observation must be made with reference to formula (38) below.

[2] The integration contour over $\epsilon$ in (15) is conveniently drawn slightly to the right of the imaginary axis, enveloping the left half-plane over an infinitely distant contour; in this case for $\text{Im}\,\epsilon < 0$ the integration contour over $t$ is shifted upward and passes through the point $2\pi i$ whereas for $\text{Im}\,\epsilon > 0$ it is shifted downward and passes through the point $-2\pi i$.

We make the linear substitution $\delta\varphi \longrightarrow \hat{S}\,\delta\varphi$ with $\det\hat{S} = 1$, which diagonalizes the matrix of the operator $S''\{\phi_c\}$ and we set

$$D\varphi = D'\varphi \prod_{i=1}^{r} d\tilde{\varphi}_i, \qquad (26)$$

where we have isolated $r$ variables of integration (denoted by the tilde) which correspond to zero eigenvalues of the operator $S''\{\phi_c\}$ and do not in fact appear in the exponential function (25). In order to ensure correct integration over zero-order modes, the following partition of unit is introduced below in the integrand (25)

$$1 = \prod_{\{i=1\}}^{r} \int d\lambda_i \delta(\lambda_i - f_i\{\varphi\}), \qquad (27)$$

where $\lambda_i$ are collective variables. An example of such a variable is the instanton center $x_0$ defined as

$$\int d^d x |\varphi(x)|^4 (x - x_0) = 0, \qquad (28)$$

for which integration of the type (27) has the form

$$1 = \int d^d x_0 \delta\left(x_0 - \frac{\int d^d x |\varphi(x)|^4 x}{\int d^d x |\varphi(x)|^4}\right). \qquad (29)$$

By introducing collective variables (which can also be the instanton "orientation," its radius, and so on [5]), we can confine ourselves to homogeneous functions $f_i\{\varphi\}$ [compare with (29)] where the degree of homogeneity can be considered to be zero without limiting the generality: if $f_i\{\mu\varphi\} = \mu^p f_i\{\varphi\}$, the substitution $\lambda_i \longrightarrow \mu^p \lambda_i$ eliminates the factor $\mu^p$ from (27). We linearize the arguments of the $\delta$-functions in (27) near the saddle-point configuration

$$1 = \prod_{i=1}^{r} \int d\lambda_i \delta(\lambda_i - f_i\{\phi_c\} - (f_i'\{\phi_c\}, \delta\varphi))$$
$$= \prod_{i=1}^{r} \int d\lambda_i \delta(\lambda_i - f_i\{\phi_c\} - \sqrt{g_c}(f_i'\{\phi_c\}, \delta\varphi)), \qquad (30)$$

and select the instanton such that $\lambda_i - f_i\{\phi_c\} = 0$ (in (28) this corresponds to a choice of solution symmetric relative to the point $x = 0$); then $\phi_c$ becomes a function of $\lambda_i$. Substituting (26) and (30) into (25) and eliminating the $\delta$-functions by integrating over $\delta\tilde{\varphi}_i$, we have

$$I_{N-1} = e^{-N} g_c^{-N+1-(M+r)/2} \det[f'\{\phi_c\}]_P$$
$$\times \int_{-\infty}^{\infty} \frac{dt}{2\pi} D'\varphi \int \prod_{i=1}^{r} d\lambda_i \phi_c^{(1)}(\lambda_i)\dots\phi_c^{(M)}(\lambda_i) \qquad (31)$$

$$\times \exp\left(-\frac{1}{2}[(\delta\varphi, S''\{\phi_c\}\delta\varphi) - Nt^2]\right),$$

where $f'\{\phi_c\}$ is an operator whose matrix consists of the columns $f_i'\{\phi_c\}$ and $[\dots]_P$ is its projection onto the subspace of the zero-order modes. The integral over $D'\varphi$ and $dt$ is determined by the determinant of the quadratic form in brackets in the exponential function (31) given by $(-N)\det[S''\{\phi_c\}]_{P'}$, although caution must be exercised when reducing this to a sum of squares [13]; the subscript $P'$ indicates a projection on a subspace complementary to the subspace of the zero-order modes. Performing elementary transformations in (31), we obtain a result having the form (3) where

$$a = \frac{1}{S\{\phi_c\}}, \quad b = \frac{M + r}{2},$$

$$c = (S\{\phi_c\})^{-(M+r)/2} \frac{(2\pi)^{(\mathcal{N}-r-2)/2} \det[f'\{\phi_c\}]_P}{\sqrt{-\det[S''\{\phi_c\}]_{P'}}} \qquad (32)$$
$$\times \int \prod_{i=1}^{r} d\lambda_i \phi_c^{(1)}(\lambda_i)\dots\phi_c^{(M)}(\lambda_i),$$

and $\mathcal{N}$ is the number of variables of integration contained within $D\varphi$ [this disappears from the answer on going over to a ratio of integrals of the type (1)].

Similar structural calculations can be made for the asymptotic form of the coefficients $A_K$. Making the substitution in (19)

$$g \longrightarrow \frac{S\{\phi_c\}}{N} g \qquad (33)$$

and isolating the dependence on $\mathcal{N}$ corresponding to the asymptotic form (3), we have

$$I_{N-1} = (S\{\phi_c\})^{-N+1} \exp(-N + N\ln N)$$
$$\times N^{(M+r-3)/2} F\left(\frac{1}{N}\right), \qquad (34)$$

where

$$F\left(\frac{1}{N}\right) = N^{-(M+r-1)/2} \oint_C \frac{dg}{2\pi i} \int D\varphi \varphi^{(1)}\dots\varphi^{(M)}$$
$$\times \exp\left(-N\frac{S\{\varphi\}}{S\{\phi_c\}g} + N - N\ln N\right)\Bigg|_{\phi = \varphi\sqrt{S\{\phi_c\}g/N}}. \qquad (35)$$

Setting $\epsilon = 1/N$ and expanding $F(\epsilon)$ as a series

$$F(\epsilon) = \tilde{A}_0 + \tilde{A}_1\epsilon + \tilde{A}_2\epsilon^2 + \dots + \tilde{A}_K\epsilon^K + \dots, \qquad (36)$$

we have by analogy with (2)

$$
\tilde{A}_{K-1} = \oint_C \frac{d\epsilon}{2\pi i} \epsilon^{(M+r-1)/2} \oint_C \frac{dg}{2\pi i} \int D\phi \varphi^{(1)} \dots \varphi^{(M)}
$$
$$
\times \exp\left(-\frac{1}{\epsilon}\left[\frac{S\{\phi\}}{S\{\phi_c\}g} - 1 + \ln g\right] - K\ln\epsilon\right)\Bigg|_{\phi = \varphi\sqrt{g\epsilon S\{\phi_c\}}}.
\tag{37}
$$

The coefficients $\tilde{A}_K$ are simply related to the unknown coefficients $A_K$ but differ from them (see below).

For large $K$ in (37) we can use the saddle-point method, for which the conditions have the form

$$
S'\{\varphi\} = 0, \quad g_c = \frac{S\{\phi\}}{S\{\phi_c\}}, \quad \epsilon_c = \frac{\ln g_c}{K},
\tag{38}
$$

and the function in the integrand for the saddle-point configuration is determined by the factor

$$
\exp(-K - K\ln\epsilon_s) \sim (\ln g_c)^{-K}K!.
\tag{39}
$$

Taking into account the substitution (33), the first two equations (38) coincide with (22) but using the solution

$$
\phi = \phi_c, \quad g_c = 1
\tag{40}
$$

leads to a singularity not to a saddle point because of (39). Thus, other solutions of the system of the first two equations (38) must be sought for which two possibilities exist.

**1. Using other branches of the logarithm.** In accordance with (38), $\epsilon_c$ is determined by the logarithm of $g_c$ and thus the substitution $g_c$ with integer $g_c \longrightarrow g_c\exp(2\pi is)$ is not an identity transformation: in this case we have $\ln g_c \longrightarrow \ln g_c + 2\pi is$. For $\phi = \phi_c$, $g_c = \exp(2\pi is)$ we have $\epsilon_c = 2\pi is/K$ and the contribution to the asymptotic form of $A_K$ is determined by the saddle points with $s = \pm 1$:

$$
A_K \sim (2\pi)^{-K}K!.
\tag{41}
$$

This is exactly the same mechanism as that used to calculate the corrections to the Stirling formula: the $g$ dependence of the function in the integrand of (19) is similar to the $x$ dependence in (11).

**2. Using other instantons.** Let us assume that $\psi_c$ is a solution of the equation $S'\{\phi\} = 0$ which differs from $\phi_c$; then on account of (38) and (39) we have the contribution to the asymptotic form

$$
A_K \sim K!\left[\ln\frac{S\{\psi_c\}}{S\{\phi_c\}}\right]^{-K},
\tag{42}
$$

which is larger the smaller $S\{\psi_c\}$. The principal contribution comes from the second instanton (see end of Section 2) and has the lower estimate

$$
A_K \gtrsim (\ln 2)^{-K}K!.
\tag{43}
$$

If $\phi_c(x)$ is a localized solution of the equation $S'\{\phi\} = 0$, we know that there also exists a solution $\psi_c(x)$ corresponding to two infinitely distant instantons $\phi_c(x)$ for which $S\{\psi_c\} = 2S\{\phi_c\}$; in general a solution $\psi_c$ can exist such that $S\{\phi_c\} < S\{\psi_c\} < 2S\{\phi_c\}$ which yields (43). Since the contribution (43) is larger than (41), in any real field theory the second of these mechanisms is the principal one; the first mechanism is only important in various nondegenerate cases such as zero-dimensional theory [see (8)] when the solution of the instanton equation is unique.

Expanding the expression in the exponential function (37) near the second instanton as far as quadratic terms in $\delta\phi$, $\delta g$, and $\delta\epsilon$ and making the substitutions $\delta g = ig_c t$, $\delta\epsilon = i\epsilon_c\tau$, we have

$$
\tilde{A}_{K-1} = g_c\epsilon_c^{-K+(r+1)/2}(S\{\psi_c\})^{-M/2}e^{-K}
$$
$$
\times \int_{-\infty}^{\infty} \frac{dt}{2\pi} \int_{-\infty}^{\infty} \frac{d\tau}{2\pi} \int D\phi \psi_c^{(1)} \psi_c^{(M)}
\tag{44}
$$
$$
\times \exp\left\{-\frac{1}{2}\left[(\delta\phi, S''\{\psi_c\}\delta\phi) - \frac{t^2}{\epsilon_c} - K\tau^2\right]\right\}.
$$

The number of zero-order modes $r'$ for the second instanton generally differs from $r$; these are isolated as before by introducing a partition of unit of the type (27), giving a dependence on $K$ having the form

$$
\tilde{A}_K \sim (\ln g_c)^{-K}\Gamma\left(K + \frac{r'-r}{2}\right).
\tag{45}
$$

In order to find the relationship between $\tilde{A}_K$ and $A_K$, we substitute (36) with $\epsilon = 1/N$ into (34), make the substitution $N \longrightarrow N+1$, and use the series reexpansion rule given in the Appendix. As a result, we obtain

$$
A_K = \frac{\tilde{A}_K}{\sqrt{2\pi}cg_c},
\tag{46}
$$

where $c$ is a coefficient appearing in the Lipatov asymptotic form (3) and determined by formula (32). Taking into account (46) and performing trivial transformations in (44), we obtain

$$
A_K = c_1\left(\ln\frac{S\{\psi_c\}}{S\{\phi_c\}}\right)^{-K}\Gamma\left(K + \frac{r'-r}{2}\right),
\tag{47}
$$

where

$$
c_1 = (S\{\psi_c\})^{-(M+r')/2}\left(\ln\frac{S\{\psi_c\}}{S\{\phi_c\}}\right)^{(r-r')/2}
$$
$$
\times \frac{(2\pi)^{(N-r'-4)/2}\det[f'\{\psi_c\}]_P}{c\sqrt{\det[S''\{\psi_c\}]_{P'}}}
\tag{48}
$$

$$\times \int \prod_{i=1}^{r'} d\lambda_i \psi_c^{(1)}(\lambda_i)\dots\psi_c^{(M)}(\lambda_i).$$

In these structural calculations, we used the form of the functional integral (1) and the homogeneity relations (18) typical of $\varphi^4$ theory; thus, the parameter $M$ appearing in the result (32) for $b$ was determined by the number of cofactors in the preexponential function (1). In other field theories, several fields of various types generally occur and the homogeneity relations differ from (18); nevertheless, for a wide range of problems the result for $b$ has the previous form (32) but the parameter $M$ has a different meaning. However, the parameter $M$ does not appear in the asymptotic formula (47), indicating that its validity is not confined to $\varphi^4$ theory: this is confirmed by the reasoning put forward in the following section.

## 4. HEURISTIC DERIVATION OF FORMULA (47)

In general, factorial series have an asymptotic form with complex parameters [see (8) and (10)] and the expansion coefficients $I_N$ are determined by the real part of some complex expression. We shall specify this for large $N$,

$$\begin{aligned}I_N = \ &\mathrm{Re}\{ca^N\Gamma(N+b)(1+\Delta_N)\\&+\tilde{c}\tilde{a}^N\Gamma(N+\tilde{b})+\dots\},\end{aligned} \quad (49)$$

taking into account the Lipatov asymptotic form $ca^N\Gamma(N+b)$, the unknown power corrections to it denoted by $\Delta_N$, and the contribution of the next instanton $\tilde{c}\tilde{a}^N\Gamma(N+\tilde{b})$; the corrections to the latter and the contribution of higher order instantons are shown by the ellipses. Removing the Lipatov asymptotic form from the brackets, we have

$$\begin{aligned}I_N = \ &\mathrm{Re}\Big\{ca^N\Gamma(N+b)\\&\times\Big[1+\Delta_N+\frac{\tilde{c}}{c}\Big(\frac{\tilde{a}}{a}\Big)^N N^{\tilde{b}-b}+\dots\Big]\Big\}.\end{aligned} \quad (50)$$

Assuming $\epsilon=1/N$, we can see that the last term has an intrinsic singularity for $\epsilon=0$ which may be attributed to the imaginary part of some factorial series [5]

$$\epsilon^{-\beta}e^{-\alpha/\epsilon}=\frac{\alpha^{-\beta}}{\pi}\mathrm{Im}\sum_K\Gamma(K+\beta)\Big(\frac{\epsilon}{\alpha}+i0\Big)^K, \quad (51)$$

whose substitution into (50) leads to an expansion in reciprocal powers of $N$. It is natural to assume that the expression in brackets (50) is an analytic function whose imaginary and real parts can only appear in a strictly determined combination. The real part of the series (51) is much larger than the imaginary one and should originate from contributions which are higher in the hierarchy than the last term in (50); only $\Delta_N$ can fulfill this role. Combining the second and third terms in brackets (50), we obtain[3]

$$\begin{aligned}I_N = \ &\mathrm{Re}\Big[ca^N\Gamma(N+b)\Big\{1+\mathrm{const}\cdot\sum_K\Gamma(K+\tilde{b}-b)\\&\times\frac{[\ln(a/\tilde{a})+i0]^K}{N^K}+\dots\Big\}\Big].\end{aligned} \quad (52)$$

The singularity on the left-hand side of (51) is associated with high-order terms of the series [see the discussion of formula (4.10) in [5]] and the form of the general term given in (52) is in fact only valid for large $K$. Bearing in mind that the parameters $a$ and $b$ of the instanton contribution have the form (32) for a wide range of field theories, we return to the result (47) where, however, the coefficient $c_1$ no longer has the specific form (48).

We shall explain the meaning of these manipulations. As we know, the expansion of the function $f(\epsilon)$ as a power series in $\epsilon$ has a radius of convergence equal to the distance between the origin and the nearest singular point $f(\epsilon)$ on the complex plane. For a factorial series the radius of convergence is zero and a singularity should be found for $\epsilon=0$. Characteristic singularities generating factorial series have the form of branch cuts at which the discontinuity decays exponentially for $|\epsilon|\longrightarrow 0$ [14] [see (51)]. It is deduced from the qualitative pattern established above that (a) $\Delta_N$ has the form of a factorial series in $1/N$; (b) the divergence of this series is determined by the second instanton; (c) the contribution of the latter in (50) contains a characteristic singularity generating these series. From this it is logical to conclude that the second and third terms in brackets (50) form a single entity, being related to the real and imaginary parts of the same analytic function.

This reasoning is merely based on the fact that the instanton contribution to the asymptotic form has the functional form (3). Thus, the result (47) is universal: it is not related to the specific field theory nor to the form of the quantity being studied (for example, single-particle or two-particle Green's function).

## 5. QUANTITATIVE RESULTS

We shall apply these results to the problem of an anharmonic oscillator [9]. This can be reduced to one-dimensional $\varphi^4$ theory [15] in which the instantons can easily be investigated (in particular, by using a mechanical analogy [16]); the localized solution of the instanton equation is unique and all other solutions are exhausted by multi-instanton configurations containing several infinitely distant instantons. Thus, as $\psi_c$ in (47) we need to take the two-instanton solution for which $S\{\psi_c\} = 2S\{\phi_c\}$, $r' = r + 1$ (an extraneous zero mode appears corresponding to the motion of two instantons relative to each other). Consequently, for an anharmonic oscillator we have

$$A_K = c_1 \left(\frac{1}{\ln 2}\right)^K \Gamma\left(K + \frac{1}{2}\right). \tag{53}$$

---

[3] The term $\Delta_N$ also contains similar contributions from higher instantons which are small compared with those contained in (52).

The dependence (53) can be compared with the results of Bender and Wu (5) using $c_1$ as the fitting parameter; results are plotted in Fig. 2 for $c_1 = -1.4$.

Higher order instantons have been little studied in multidimensional $\varphi^4$ theory. An exception is the four-dimensional case for which an infinite series of instanton solutions was obtained analytically by Ushveridze [17]. The second instanton in this series (following the Lipatov one ($S\{\phi_c\} = 16\pi^2/3$) has the action $S\{\psi_c\} = 9\pi^2$ which gives the result for the asymptotic form of $A_K$

$$A_K = c_1 \left(\ln\frac{27}{16}\right)^{-K} \Gamma\left(K + \frac{3}{2}\right) \tag{54}$$

(we assumed that $r' = r + 3$ because in view of the absence of spherical symmetry for the second instanton, three zero modes are added corresponding to its rotations in four-dimensional space). Unfortunately, there is no evidence that the Ushveridze series exhausts all the solutions; thus, the result (54) should be understood as a preliminary or lower estimate.

A method of determining the complete series of higher order instantons numerically was proposed in [18]. It would be desirable to use this method to check the result (54) and to find the second instantons in all existing field theories.

## 6. CONCLUSIONS

Expression (4) can be used to interpolate the coefficient function, by truncating the series at a finite number of terms and selecting the parameters $A_K$ to ensure agreement with the lowest orders of perturbation theory known from direct diagrammatic calculations. This procedure is highly acceptable and can reliably estimate the error but nevertheless is unsatisfactory in many respects. This is because when diverging series are summed, the analytic properties of the coefficient function [19] are significant and these are reproduced quite incorrectly in this procedure: the coefficient function is assigned a multiple pole at $N = 0$ but the intrinsic singularity is lost at $N = \infty$ because of the factorial divergence of the series in (4) [see formula (51)]. Qualitative allowance for the functional form of the asymptotic $A_K$ in the form $c_1 a_1^K \Gamma(K + b_1)$ enables us to select basis functions exhibiting correct behavior for $N \longrightarrow \infty$ which should give a positive effect even when the number of fitting parameters is constant. Quantitative calcu-

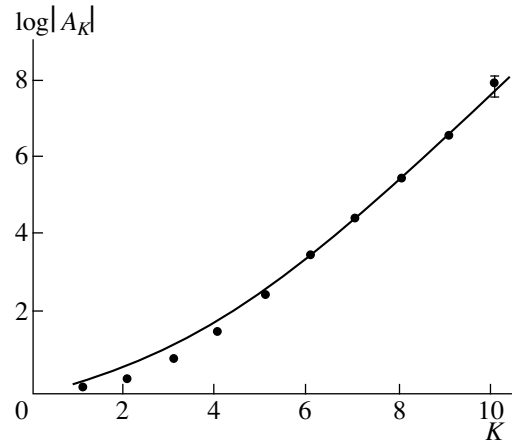

**Fig. 2.** Comparison between the asymptotic formula (53) for $c_1 = -1.4$ and the coefficients $A_K$ determined numerically in [9]. The value of $|A_{10}|$ is given in [9] with a single significant digit ($1 \times 10^8$) and the error corresponding to the range $(0.5–1.5) \times 10^8$ is indicated in the figure.

lations of the asymptotic form can be used to determine three parameters $a_1$, $b_1$, and $c_1$ characterizing the coefficient function which is equivalent in efficiency to calculating the next three orders of perturbation theory. The calculations of $a_1$ and $b_1$ do not require functional integrals and can be reduced to solving nonlinear differential equations: the calculations of $c_1$ are of approximately the same complexity as the calculations of the leading Lipatov asymptotic form. This is incomparably easier than calculating the successive terms of a perturbation theory series where progression to the next order takes, on average, ten years

The author thanks L.N. Lipatov for discussions of preliminary results of this work and participants at seminars at the Institute of Physical Problems and the Physics Institute of the Russian Academy of Sciences for their interest in this work and useful discussions.

## ACKNOWLEDGMENTS

This work was supported by INTAS (grant no. 96-0580) and the Russian Foundation for Basic Research (project no. 00-02-17129).

*APPENDIX*

*Derivation of Formula (8)*

Let us assume that two expansions exist:

$$F_N = 1 + \frac{A_1}{N} + \frac{A_2}{N^2} + \ldots + \frac{A_K}{N^K} + \ldots, \tag{A.1}$$

$$F_N = 1 + \frac{B_1}{N+p} + \frac{B_2}{(N+p)^2} + \ldots + \frac{B_K}{(N+p)^K} + \ldots \tag{A.2}$$

If the second series is factorial,

$$B_K = \text{Re}\{ca^K \Gamma(K+b)\}, \quad K \longrightarrow \infty, \quad (A.3)$$

it is easy to show by direct reexpansion that

$$A_K = \text{Re}\, e^{-p/a} ca^K \Gamma(K+b), \quad K \longrightarrow \infty. \quad (A.4)$$

Making the substitution $N \longrightarrow N + \beta - 1$ in (9) and using (A.3) and (A.4), we obtain the result

$$\Gamma(N+\beta) = N^{\beta-1}\sqrt{2\pi N}e^{-N}N^N$$

$$\times \left\{ 1 + \ldots - \frac{2\Gamma(K)}{N^K}\text{Re}\frac{e^{-2\pi i\beta}}{(2\pi i)^{K+1}} + \ldots \right\} \quad (A.5)$$

and, applying it to relation (7) and using factorial series algebra [5], derive formula (8).

*Translation was provided by AIP*